\documentclass[twocolumn,showpacs,preprintnumbers,amsmath,showkeys, amssymb]{revtex4}
\usepackage[unicode]{hyperref}
\usepackage[caption=false]{subfig}
\usepackage{graphics}
\usepackage{epsfig}
\usepackage{graphicx}
\graphicspath{ {./images/} }
\usepackage{dcolumn}
\usepackage{bm}
\usepackage[caption=false]{subfig}
\usepackage[utf8]{inputenc}
\usepackage{dirtytalk}
\usepackage[autostyle]{csquotes}
\usepackage{xcolor}
\usepackage{scalerel}

\begin{document}

\title{
Multi-multifractality and dynamic scaling in stochastic porous lattice
}

\author{Tushar Mitra and Md. Kamrul Hassan
}
\date{\today}

\affiliation{
Theoretical Physics Group, Department of Physics, University of Dhaka, Dhaka 1000, Bangladesh
}

\begin{abstract}%

In this article, we extend the idea of stochastic dyadic Cantor set to weighted planar stochastic 
lattice that leads to a stochastic porous lattice. The process starts with an initiator which
we choose to be a square of unit area for convenience. We then define a generator that
divides the initiator or one of the blocks, picked preferentially with respect to their areas,
to divide it either horizontally or vertically into two rectangles of which 
one of them is removed with probability $q=1-p$. We find that the remaining number of blocks and 
their mass varies with time as $t^{p}$ and $t^{-q}$ respectively. Analytical solution shows that the 
dynamics of this process is governed by infinitely many hidden conserved quantities each
of which is a multifractal measure with porous structure as it contains missing blocks of 
various different sizes. The support where these measures are distributed is fractal with fractal 
dimension $2p$ provided $0<p<1$. We find that if the remaining blocks are characterized 
by their respective area then the corresponding block size distribution function obeys dynamic scaling.

\end{abstract}

\pacs{61.43.Hv, 64.60.Ht, 68.03.Fg, 82.70Dd}

\keywords{Planar lattice, Cantor set, Fractal, Multifractal, Dynamic scaling, Self-similarity}

\maketitle

\section{Introduction}

We live in an Euclidean space of dimension three and the geometry of objects that we see or interact with are also considered to be Euclidean. Geometry has always been an exceptional tool to describe natural and man-made objects using lines, squares, circles etc. However, not all objects are Euclidean. Rather
 there are objects in nature that are accompanied by such an extent of complexity that they 
 cannot be described by Euclidean geometry and till 1975 they were called geometric
 monsters. Benoit B. Mandelbrot conceived, introduced and disseminated the idea of fractal to describe these geometrically complex objects \cite{ref.mandelbrot1, ref.mandelbrot2}. It has revolutionized
 the idea of geometry and has been found useful in almost every branch of academia. There is no
 well defined and well accepted definition of fractal. Typically, an object that have
 non-integer dimension and have self-similar property is said to be a fractal. However,
 recently we have defined it as an object whose dimension $d_f$ is less than that of its embedding
 space $d$. In the case, when $d=d_f$ the constituents of the object is uniformly distributed 
 and as $d_f<d$ it guarantees that the density of constituents is not constant rather decays
 following a power-law with the linear system size $L$. It also implies that the smaller the
 value of $d_f$ in comparison to $d$, the higher the inhomogeneity of the constituents. 
  Natural objects like broccoli, snowflake, coastline of an island, thunder etc. can be best described as fractals albeit they are not strictly self-similar rather they are  
  statistically self-similar. Fractals have been relevant in various subjects like physics, chemistry, biology, earth science, economics, arts etc.  \cite{ref.edgar, ref.korvin, ref.vicsek}. Recently a group of researchers have created electronic quantum fractal where electron wave functions are delocalized over the Sierpiński structure \cite{ref.qfractal}.

One of the simplest text-book example of fractal is definitely the triadic Cantor set. 
The process starts with an initiator of closed interval [0,1]. A generator divides it 
into three equal parts and deletes the open middle third. The generator is then applied 
on the remaining intervals \textit{ad infinitum}. The resulting set has a non-integer 
Hausdorff-Bessicovitch dimension $d_f=\log 2/ \log 3$ which is less than that of the 
dimension of the space $d=1$ where the set is embedded. The Cantor set problem has been 
relevant in many pedagogical and practical purposes \cite{ref.cantor1, ref.cantor2}. 
However, the simple construction of Cantor set lacks in two ways. Firstly, it does not 
appear through evolution in time. Secondly, it is not governed by any sort of randomness 
throughout its construction process. This two properties are the most common features of 
natural fractals. To overcome these Hassan et al. proposed different variants of the 
Cantor set like dyadic Cantor set, kinetic and stochastic dyadic Cantor set including 
randomness and evolution with time, which has been useful in many disparate systems 
\cite{ref.pandit}. Recently one of us have shown that the conservation law of stochastic 
fractals and symmetry is in accordance with Noether's theorem \cite{ref.special, ref.shahnoor}.

Constituents that make objects are not always distributed uniformly like we see in
Euclidean and fractal objects. In reality, many quantities fluctuates wildly. For instance, gold, diamond
and many other precious materials are not found at equal concentration everywhere on the Earth.
Rather it is found in high concentrations at only a few places, in low concentrations at many places, 
and in very low concentration at almost everywhere on Earth. 
To measure the fractal dimension of an object that constitute a set
$S$ we usually partition the embedding space into hyper-cubic cells with 
size of their side $\delta$. We then count the number of hyper-cubes $N(\delta)$ that contain at 
least one point of the set $S$, varying the size of $\delta$ and measuring the corresponding $N(\delta)$. Slope of the straight line obtained by plotting these data in the $\log-\log$ scale gives the fractal dimension. The important thing that is missing no matter how many points of the set each cell
contain as long as they contain at least one point all are counted with equal weight.
That is, this method does not give any information about detailed nature of the distribution
and hence it is the crudest form of measure. The question is: Is there a way to give a higher weight 
to hyper-cubic boxes with higher population of points and a lower weight to those which contain
lower population of points? Multifractal formalism is an answer to this question. In 2010
Hassan {\it et al.} proposed a weighted planar stochastic lattice(WPSL) which has been
shown to have many non-trivial properties \cite{ref.hassan_njp, ref.hassan_jpc}. One of the properties 
is that it is not only a multifractal but infinitely many multifractal spectrum. A multifractal 
system is a generalization of a fractal system where a single exponent (the fractal dimension) 
is not enough to describe the system, rather a continuous spectrum of exponents are needed \cite{ref.multifractal}. Multifractality has been observed in events like Anderson localization, quantum hall transition and has been useful in many different fields that includes turbulence in fluid dynamics, voltage and current distribution in random resistor networks, heartbeat dynamics, growth by diffusion-limited aggregation (DLA), collision cascade to name a few  \cite{ref.anderson, ref.hall, ref.ivanov}. Recently, we have proposed another close variant of WPSL that we named as WPSL1. The lattice provides different properties than that of previously proposed WPSL and is a small-world \cite{ref.tushar_hassan}.

In this article, we extend the idea of stochastic dyadic Cantor set to two dimensional case and see how it 
differs from its one dimensional counterpart. We start the process with a square of unit area. 
To apply on this initiator over and over again we define a generator that
divides a block either horizontally or vertically at random into two smaller block and remove one of them
with probability $1-p$. It is then applied to only block at each time step after picking it 
preferentially according its area. Due to removal of blocks we actually create structure that 
can be treated as porous media since it contains pores or void of different sizes due to removal of mass. 
We can use such porous structure as a backbone while solving theoretical models to gain deeper 
insight into various physical phenomena such as percolation. It can also be considered
as fragmentation of two dimensional particles with mass loss. It has many interesting and non-trivial properties.
We show that it is governed by infinitely many conservation laws even though the conservation
of mass law is violated. It is shown that each of these conserved quantity can be regarded
as multifractal measure and hence the structure is not only a multifractal rather infinitely many multifractal. It has been shown that if the blocks of the structure is characterized by their respective
areas then the size distribution function obeys dynamic scaling and this is verified
by using the idea of data-collapse. It suggests that the structure is self-similar in time.

The organization of the rest of this paper is as follows. In section II, the connection between stochastic dyadic Cantor set and weighted planar stochastic lattice is developed. In section III, we propose a porous stochastic lattice by incorporating maass loss in WPSL1 and describe the algorithm. In section IV, various geometric properties of the stochastic porous lattice are discussed and we show how remaining area and number of blocks varies with time $t$. The conservation laws are also discussed in the section. In section V, multifractal analysis of the lattice is performed and we show that it is a infinitely many multifractal. In section VI, the area distribution function is introduced and we show that the lattice possesses temporal self-similarity. Finally, results are discussed and conclusions drawn in section VII.

\section{From dyadic Cantor set to weighted planar stochastic lattice}

Weighted planar stochastic lattice was first proposed in 2010 by Hassan et al. which provides many non-trivial topological and geometric properties \cite{ref.hassan_njp}. The substrate is 
typically chosen as a square of unit area and is called an initiator. In step one, 
a point on the substrate is chosen at random and two orthogonal lines parallel to its sides are grown to
divide the substrate randomly into four smaller blocks. In the next step and all
the steps thereafter one of the blocks is first picked with probability of their respective area and the generator
divides that randomly into four smaller blocks like it is done in step one to the initiator.  
The resulting lattice is called weighted planar stochastic lattice or in short WPSL2 
where the factor $2$ refers to the fact that two mutually perpendicular partitioning lines 
are used. It has much richer properties than the square lattice and kinetic square lattice \cite{ref.hassan_njp,ref.hassan_jpc}. One of the emergent behavior of the lattice is that the coordination number distribution function of WPSL2 follows inverse power-law which immediately implies that the degree distribution of the network 
corresponding to its dual follow the same inverse power-law $P(k)\sim k^{-\gamma}$ with the same exponent $\gamma=5.66$ \cite{ref.hassan_njp}.
They have shown that it is governed by infinitely many conservation laws and one of conserved
quantity can be used as multifractal measure and geometrically the lattice is a multifractal. Later it was shown that each of the infinitely many conservation laws is actually a multifractal measure and hence WPSL2 is a multi-multifractal \cite{ref.dayeen_csf}.

Recently, we have studied yet another variant of WPSL where a block chosen preferentially with respect to area is divided into two blocks 
by dividing it either horizontally or vertically with equal {\it a priori} probability. The resulting lattice is called WPSL1, where the factor 1 represents the fact that at each time step only one partitioning line is grown rather than two which was the case for WPSL2. We show WPSL1 too provides awe-inspiring topological and geometric properties. For instance, the degree distribution again follow a power law 
$P(k) \sim k^{-\gamma}$ with $\gamma = 4.13$ which is much closer to real life scale-free network. 
It has also been found that it has small-world and nested hierarchical network properties. Geometrically WPSL1 also has infinitely many a multi-multifractal spectrum for each of the infinitely many non-trivial conserved quantities although trivial area conservation do not give rise to multifractal spectrum \cite{ref.tushar_hassan}.

We are particularly interested in WPSL1. If we compare it with the stochastic dyadic Cantor set(SDCS), we can see that WPSL1 is exactly the two dimensional version of the stochastic dyadic Cantor set, except for the fact that no cell (interval for the case of SDCS) is removed from the initiator in each time step. So, the support of WPSL1 remains an Euclidean two dimensional space, although WPSL1 itself is regarded as a multi-multifractal. We now want to extend WPSL1 to a new model where there is probabilistic mass loss involved, which may introduce a fractal support and can mimic the SDCS in higher dimension.

\section{Construction of stochastic porous lattice}

\begin{figure}
\centering

\includegraphics[width=8.5cm,height=8.0cm,clip=true]
{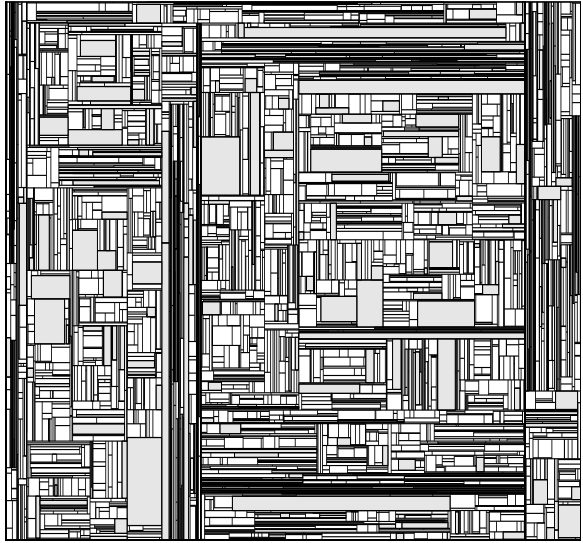}
\label{fig:2}

\caption{A snapshot of the stochastic for p=0.95 and t=5000. The shaded cells indicate that the cells were deleted.
}

\label{fig:2}
\end{figure}

There exist a great extent of demand for a structure that can mimic the porous media. That is, 
structure that contain voids or gaps. To this end we now modify the construction process of WPSL1 
so that at each time step of its construction process one of the smaller blocks will be
removed with a probability $p$ of our choice. 
In step one, the generator divides the initiator randomly into two smaller blocks either horizontally or vertically. 
We then label the left block by $a_1$ if it is divided horizontally and the top block
by $a_1$ if it is divided vertically. The other block is always labelled 
as $a_2$. The second block $a_2$ is then kept with a probability $p$ and removed with probability $1-p$. In each step thereafter
only one block is picked preferentially with respect to their respective area 
 and then it is divided in the same way as has been done to the initiator. 
 In general, the $j$th step of the algorithm can be described as follows.
\begin{itemize}
\item[(i)] Subdivide the interval $[0,\sum_{i=1}^{l} a_i]$ into $l$ sub-intervals of size $[0,a_1]$, $[a_1, a_1+a_2]$,$\  ...$, 
$[\sum_{i=1}^{l-1} a_i,\sum_{i=1}^{l} a_i]$ each of which represents the blocks labeled by their areas $a_1,a_2,...,a_{l}$ respectively. Here we have assumed that there are $l$ number of remaining blocks in the $j^{th}$ step. Then, the area sub-intervals of the deleted cells or voids fall within the interval $[\sum_{i=1}^{l} a_i,1]$.
\item[(ii)] Generate a random number $R$ from the interval $[0,1]$ and find which of the $l$ sub-interval 
contains this $R$. The corresponding block it represents, say the $k$th block of area $a_k$, is picked. If $R$ falls within any of the voids left by removal, then increase time by one unit and go back to step (i). 
\item[(iii)] Calculate the length $x_k$ and the width $y_k$ of this block and keep note of the coordinate of the
lower-left corner of the $k$th block, say it is $(x_{low}, y_{low})$.
\item[(iv)] Generate two random numbers $x_R$ and $y_R$ from $[0,x_k]$ and $[0,y_k]$ respectively and hence
the point $(x_{R}+x_{low},y_{R}+y_{low})$ mimics a random point chosen in the block $k$.
\item[(v)] Generate a random number $p$ within $[0,1]$. 
\item[(vi)] If $p<0.5$ then draw a vertical line else a horizontal line through the point $(x_{R}+x_{low},y_{R}+y_{low})$ 
 to divide it into two smaller rectangular blocks. The label $a_k$ is now redundant and hence
it can be reused.
\item[(vii)] Label the left or top of the two newly created blocks as $a_{l}$ depending
whether a vertical or horizontal line is drawn respectively and the remaining block is then labeled as $a_{l+1}$.
\item[(viii)] Generate another random number $R_1$ within $[0,1]$. If $R_1<p$ then the block $a_{l+1}$ is kept and if $R_1\geq p$, the block is removed.
\item[(ix)] Increase time by one unit and repeat the steps (i) - (viii) {\it ad infinitum}.
\end{itemize}
 
We name the resulting lattice as stochastic porous lattice. A snapshot of the lattice is shown Fig. (\ref{fig:2}), which reveals the intriguing pattern of surviving blocks and voids. The number of blocks $N(t)$ for the stochastic porous lattice grows with time $t$ but $N(t)$ now becomes probabilistic rather deterministic. The trivial conservation of total area also is no longer applicable for the porous lattice and it too, depends on time $t$.

\section{Geometric properties of stochastic porous lattice}

In this section we shall try to solve some aspects of the model analytically. To this
end, we  characterize the remaining blocks of the system by their length $x$ and width $y$. We
then define the block size distribution function $f(x,y,t)$ such that
$f(x,y,t)dxdy$ is the number of blocks whose length and width fall within the range 
$[x,x+dx]$ and $[y,y+dy]$. Then  
the evolution of their distribution function $f(x,y,t)$ can be described by the 
master equation
\begin{eqnarray}
\label{eq:WPSL}
{{\partial f(x,y,t)}\over{\partial t}} & = & -xy f(x,y,t)+\dfrac{1+p}{2}x\int_y^\infty f(x,y_1,t)dy_1  \nonumber \\ & +
&  \dfrac{1+p}{2}y\int_x^\infty f(x_1,y,t)dx_1.
\end{eqnarray}
The first term on the right accounts for the loss of blocks of sides $x$ and $y$ due to nucleation of seed
of crack on to it. The pre-factor $xy$ here implies that the blocks are picked 
preferentially according to their area.
The second term on the right account for the gain of block of sides $x$ and $y$ upon breaking
the block of sides $x$ and $y_1>y$ horizontally with probability 1/2. The factor $(1+p)$ stands for the fact that on average $1+p$ number of blocks are created with each breaking event. The last term similarly accounts for the contribution due to breaking  a block of sides $x_1>x$ and $y$ vertically.

Solving  Eq. (\ref{eq:WPSL}) for $f(x,y,t)$ with any given initial size distribution $f(x,y,0)$ is a 
formidable task. Therefore, instead of solving for $f(x,y,t)$ we find
it instructive to solve it for $2$-tuple Mellin transform of $f(x,y,t)$ given by
\begin{equation}
\label{momenteq_1}
{M(m,n;t)  =  \int_0^\infty \int_0^\infty x^{m-1} y^{n-1} f(x,y,t) dx dy.
}
\end{equation}
Incorporating it in Eq. (\ref{eq:WPSL}) leads us to,
\begin{equation}
\label{momenteq_2}
{{dM(m,n;t)}\over{dt}} = - {{2mn-b(m+n)}\over{2mn}}M(m+1,n+1;t),
\end{equation}
where $b=(1+p)$. It can be seen that if we choose $m=m^\dagger$ and
\begin{equation}
    n={{bm^\dagger}\over{2m^\dagger -b}},
\end{equation}
then the corresponding moment is a conserved quantity i.e. 
\begin{equation}
\label{momenteq_const}
M\Big(m^\dagger, {{m^\dagger(1+p)}\over{2m^\dagger-(1+p)}};t\Big)=const., 
\end{equation}
$\ \forall \ m^\dagger$ and $\forall\  0<p<1$. Thus, the system is governed by infinitely
many non-trivial conserved quantities.
The credit for such striking feature goes to analytical approach to the problem without which would not
have been possible to even guess that such infinitely many conservation law exist. Unlike
WPSL1 and WPSL2, the present model does not obey conservation of mass law 
which is obvious since mass is removed during the process with a non-zero probability.

\begin{table}
\centering
\setlength{\tabcolsep}{20pt}
\begin{tabular}{c|c|c}
\hline
p & M(1,1;t) & M(2,2;t) \\
\hline
$0.25$ & $1.10326 t^{0.25}$ & $1.10326 t^{-0.75}$ \\ 
$0.50$ & $1.12838 t^{0.50}$ & $1.12838 t^{-0.50}$ \\
$0.75$ & $1.08807 t^{0.75}$ & $1.08807 t^{-0.25}$ \\
\end{tabular}
\caption{Exact values of the moments as a function of time.}
\label{tab:table1}
\end{table}

To find the solution for $M(m,n;t)$ we find it convenient to re-write  Eq. (\ref{momenteq_2}) 
as 
\begin{equation}
\label{momenteq_3}
{
{{dM(m,n;t)}\over{dt}} = - \Big ( {\alpha_{+}\alpha_{-}\over{m n}}\Big )M(m+1,n+1;t).
}
\end{equation}
with,
\begin{equation}
\label{momenteq_4}
{
\alpha_{\pm}={{m+n-b}\over{2}}\pm\sqrt{ {{(m-n)^2}\over{4}}+{{b^2}\over{4}}}.
}
\end{equation}
On the other hand, the Taylor series expansion of M(m,n;t) about $t = 0$ is
\begin{equation}
    \label{eq:taylor}
    M(m,n;t)=M(m,n,0)+\sum_{j=1}^\infty {{t^j}\over{j!}}{{d^jM(m,n;t)}\over{dt^j}}\Big |_{t=0}.
\end{equation}
We now iterate Eq. (\ref{momenteq_3}) to find various derivatives of $M(m,n;t)$ which are given by,
\begin{eqnarray}
\label{momenteq_5}
\dfrac {d^{j}M(m,n;t)} {dt^{j}} = (-1)^{j} \prod_{k = 1}^{j} \dfrac {(\alpha_{+}+k-1)(\alpha_{-}+k-1)} {(m+k-1) (n+k-1)} \nonumber \\ M(m+j,n+j;t). \nonumber \\
\end{eqnarray}
Using mono-disperse initial condition 
\begin{equation}
f(x,y,0)=\delta(x-1)\delta(y-1),    
\end{equation} 
gives $M(m,n;0)$ and all its derivatives equal to one. Substituting all these in Eq. (\ref{eq:taylor})
gives a solution for $M(m,n;t)$ as  
\begin{equation}
\label{momenteq_6}
M(m,n;t) =  ~_{2}F_{2} (\alpha_{+},\alpha_{-};m,n;-t),
\end{equation}
where $~_{2}F_{2}(a,b;c,d;z)$ is known as generalized hypergeometric function \cite{ref.hypergeometric}.
 
 Using the asymptotic behavior of generalized hypergeometric function we find that the moment $M(m,n;t)$ decays algebraically with time
 \begin{equation}
\label{momenteq_7}
M(m,n;t) =  \dfrac {\Gamma(m) \Gamma(n) \Gamma(\alpha_{+}-\alpha_{-})} {\Gamma(\alpha_{+})\Gamma(m-\alpha_{-})\Gamma(n-\alpha_{-})} t^{-\alpha_{-}},
\end{equation}
Using it we immediately find that the first moment that represents the number of intervals increases with 
time as
\begin{equation}
\label{momenteq_nc}
M(1,1;t) \sim t^{p} ,
\end{equation}
and the second moment that represents total mass decays with time as
\begin{equation}
\label{momenteq_ac}
M(2,2;t) \sim t^{-(1-p)},
\end{equation}
as expected since the model describes WPSL1 with mass loss.
It can be seen from the above equations the total number of blocks grows following a power law with time with exponent $p$ and total area is not a conserved quantity, rather it decays inversely with time with an exponent $1-p$. The exact values of $M(1,1;t)$ and $M(2,2;t)$ for different $p$ are shown in Table (\ref{tab:table1}) as a function of time $t$ calculated using Eq. (\ref{momenteq_7}).

\begin{figure}
\centering

\includegraphics[width=8.5cm,height=8.0cm,clip=true]
{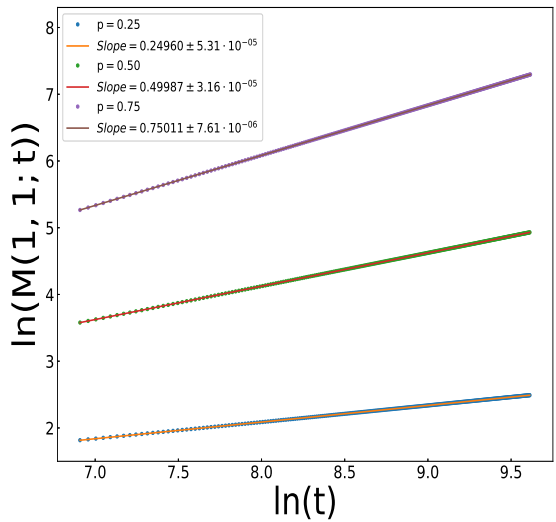}
\label{fig:N}

\caption{ Plot of $ln[(M(1,1;t)]$ vs $ln(t)$ for $p=0.25$, $p=0.5$ and $p=0.75$. The resulting plots are clearly straight line with slopes exactly equal to $p$ for all three cases.
}

\label{fig:N}
\end{figure}

 We have performed extensive numerical simulations to verify our results.
 First of all, the results based on numerical simulation 
 for $M(1,1;t)$ and $M(2,2;t)$ are shown in Fig. (\ref{fig:N}) and Fig. (\ref{fig:area}) respectively 
 for $p = 0.25, 0.5$ and $0.75$ in both cases. All the results are extracted from a large number of independent realizations. It can be seen from the plots that the results matches perfectly with the analytical prediction given by Eqs. (\ref{momenteq_nc}) and (\ref{momenteq_ac}). Moreover, the fitted straight line in Fig. (\ref{fig:N}) follows $M(1,1;t)=1.10428t^{0.24961}$ for $p=0.25$, $M(1,1;t)=1.13006t^{0.49987}$ for $p=0.5$, and $M(1,1;t)=1.08655t^{0.75011}$ for $p=0.75$ which matches perfectly with the calculated values given in  Table (\ref{tab:table1}). On the other hand, the fitted straight line in Fig. (\ref{fig:area}) follows $M(2,2;t)=1.10643t^{-0.74979}$ for $p=0.25$, $M(2,2;t)=1.12731t^{-0.49961}$ for $p=0.5$, and $M(2,2;t)=1.08546t^{-0.24979}$ for $p=0.75$ which also matches perfectly.  This shows the high accuracy and precision between the theory and numerical simulation.

\begin{figure}
\centering

\includegraphics[width=8.5cm,height=8.0cm,clip=true]
{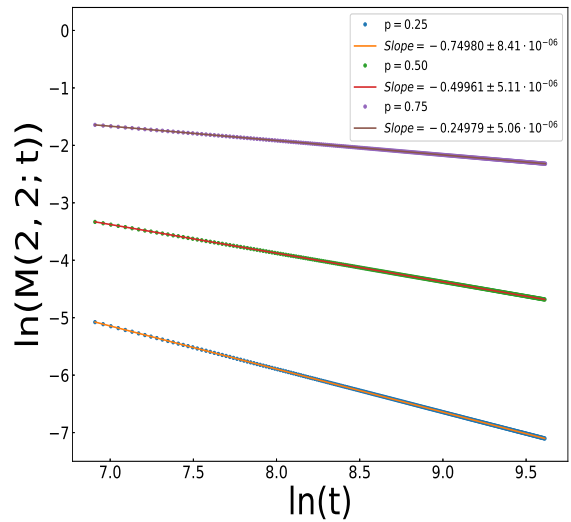}
\label{fig:area}

\caption{ Plot of $ln[(M(2,2;t)]$ vs $ln(t)$ for $p=0.25$, $p=0.5$ and $p=0.75$. The resulting plots are clearly straight lines with slope exactly equal to $-(1-p)$ for all three cases.
}

\label{fig:area}
\end{figure}

\begin{figure}
\centering

\includegraphics[width=8.5cm,height=8.0cm,clip=true]
{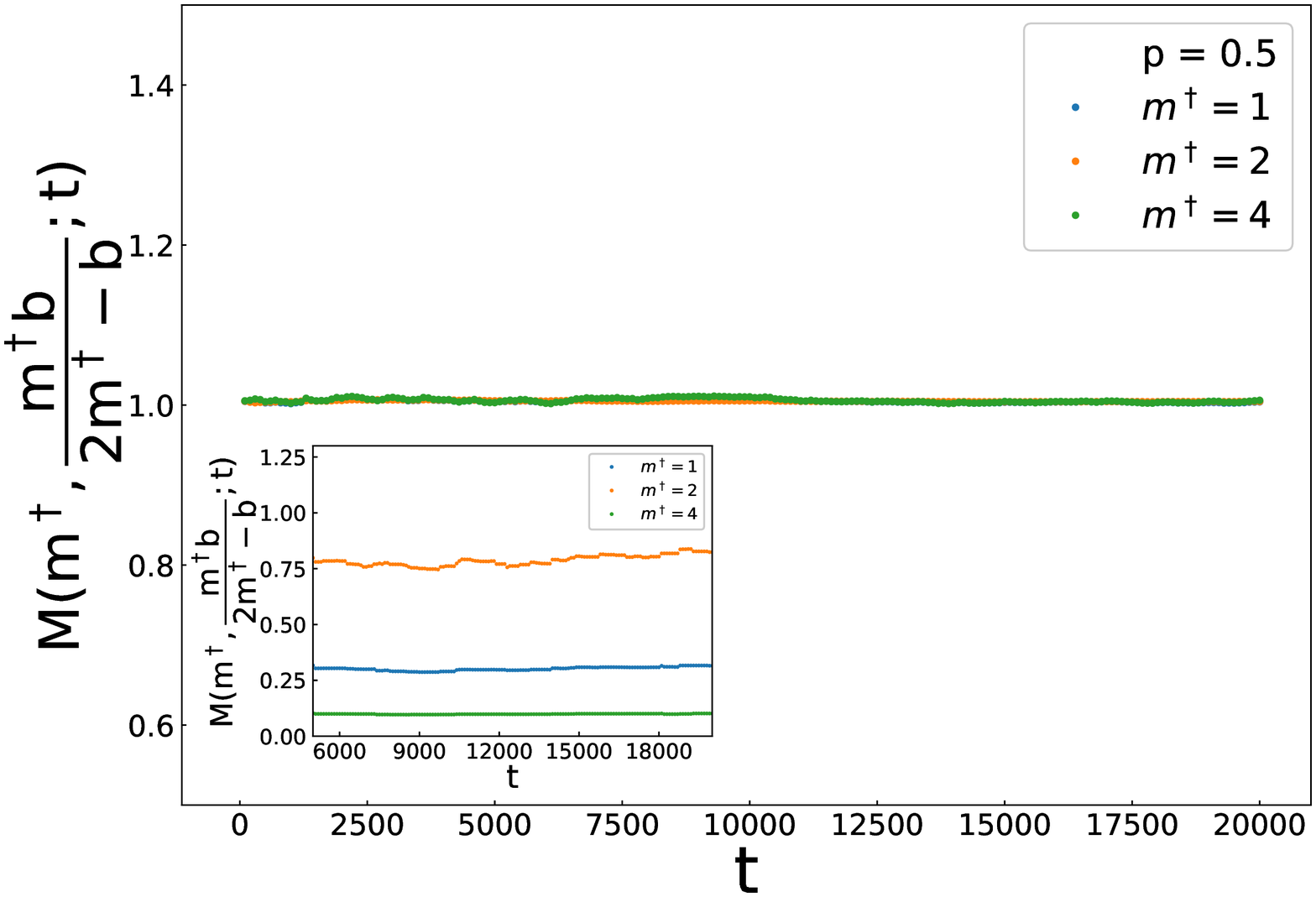}
\label{fig:4}

\caption{ Plots of the moment $M(m^\dagger,{{m^\dagger b}\over{2m^\dagger-b}};t)$ vs $t$ for $p=0.5$ and $m^\dagger = 1, 2$ and $4$ for an average of $20000$ independent realizations. It is evident from the plot that the moments are constant with time and equal to $1$ for all values of $m^\dagger$. In the inset the same plot for a single realization where a considerable number of remaining blocks are present in the lattice is shown. It reveals that although $M(m^\dagger,{{m^\dagger b}\over{2m^\dagger-b}};t)$ remains constant with time $t$ for all $m^\dagger$, the constant values are different from one another for a single realization.
}

\label{fig:4}
\end{figure}

To verify the conservation laws given by Eq. (\ref{momenteq_const}) we first write its discrete counterpart
\begin{equation}
\label{consteq}
\sum_i^N x_i^{m^\dagger -1} y_i^{ {{m^\dagger b}\over{2m^\dagger-b}}-1}=const.
\end{equation}
To this end we plot $M(m^\dagger,{{m^\dagger b}\over{2m^\dagger-b}};t)$ using Eq.(\ref{consteq}) as a function of $t$ in the inset of Fig. (\ref{fig:4}) for $p=0.5$ $m^\dagger = 1, 2$ and $4$ for a single realization where there are a considerable number of remaining blocks. The moment for each $m^\dagger$ value remains constant with time albeit the value of the moments are different for different values of $m^\dagger$. In fact, the value of the moment for a fixed value of $m^\dagger$ is different in every realization, although it remains constant with time in every case.
In addition, for $m=m^\dagger$ and $n={{bm^\dagger}\over{2m^\dagger -b}}$ the coefficient of $t^{-\alpha_{-}}$ and $\alpha_{-}$ in Eq.(\ref{momenteq_7})  after some calculation turns out to be equal to $1$. This implies that,
\begin{equation}
\label{moment_final}
M\Big(m^\dagger,{{m^\dagger b}\over{2m^\dagger-b}};t\Big) = 1  \ \forall \ m^\dagger > 0, t > 0, p > 0
\end{equation}
The resulting data for $M(m^\dagger,{{m^\dagger b}\over{2m^\dagger-b}};t)$ as a function of $t$ is plotted in Fig. (\ref{fig:4}) that is averaged over $20000$ independent realizations for $p=0.5$ and $m^\dagger=1, 2$ and $4$. The plot shows that the conserved quantity is equal to 1 independent of the value of $m^\dagger$ and $t$, which confirms our theoretical prediction. The moment for $p=0.25$ and $p=0.75$ were also checked and found to behave same as it is for $p=0.5$. So, the conserved quantity is always equal to 1 for every $p$, $m^\dagger$ and $t$.

\section{Multifractal analysis of stochastic porous lattice}

In this section, we show that each of the non-trivial conserved quantity can be used as a multifractal 
measure.
We assume that $\mu_i=x_i^{m^\dagger-1} y_i^{ {{m^\dagger b}\over{2m^\dagger-b}}-1}$ is the 
fraction of the total measure $\sum_i^N x_i^{m^\dagger-1} y_i^{ {{m^\dagger b}\over{2m^\dagger-b}}-1}$ 
which is found in the $i$th block of the remaining blocks. We can now  
construct the partition function 
\begin{equation}
Z_q=\sum_i\mu_i^q    
\end{equation} 
of the probability $\mu_i$. Comparing it with the definition of the two-tuple Mellin transform  
of $f(x,y,t)$ we can write it as  
\begin{equation}
\label{partitioneq}
Z_q=M \Bigg( (m^\dagger-1)q+1, \Big({{m^\dagger b}\over{2m^\dagger-b}}-1\Big)q+1;t\Bigg).
\end{equation}
According to Eq. (\ref{momenteq_7}) we, therefore, have the asymptotic solution for the partition function 
\begin{eqnarray}
\label{partitioneq1_2}
Z_q(t)\sim t^{\scaleobj{0.8}{\dfrac{\sqrt{(m^\dagger-{{{m^\dagger b}\over{2m^\dagger-b}})^2q^2+b^2}}-\Big((m^\dagger+{{{m^\dagger b}\over{2m^\dagger-b}}-2)q+2-b \Big)}}{2}}}
\end{eqnarray} 
We now measure $Z_q$ using the square root of the mean area,
\begin{equation}
\label{delta}
\delta(t) = \sqrt{\dfrac{M(2,2;t)}{M(1,1;t)}} \sim t^{-1/2},
\end{equation}
as the yard-stick and find that it decays following a power-law
\begin{equation}
\label{weightednumber}
Z_q(\delta)\sim \delta^{-\tau(q)},
\end{equation}
where the mass exponent 
\begin{eqnarray}
\label{massexponent}
\tau(q) & = & \sqrt{(m^\dagger-{{{m^\dagger b}\over{2m^\dagger-b}})^2q^2+b^2}} - \nonumber\\ & 
 & \Big((m^\dagger+{{{m^\dagger b}\over{2m^\dagger-b}}-2)q+2-b\Big)}.
\end{eqnarray} 
The mass exponent must possess the properties such that $\tau(0)$ is the dimension of the support and $\tau(1)=0$ required by the normalization of the probabilities $p_i$. The later one is clearly true in our case since $\tau(1)=0$. On the other hand, $\tau(0)=2p$, which reveals that the dimension of the
support is not an Euclidean as long as $p<1$. Rather it is a fractal with fractal dimension $2p$
which is always less than the dimension of the space $d=2$ where the lattice is embedded for $p<1$. 
This is shown in Fig. (\ref{fig:tau}), where $\tau(q)$ is plotted as a function of $q$ for $p=0.25$
 and $m^\dagger=5, 6$ and $7$. It is clear from the plot that $\tau(0)=2p$, in this particular case
 $\tau(0)=0.5$ for any value of $m^\dagger$. Also, $\tau(1)=0$ for any value of $m^\dagger$ and $p$.

\begin{figure}
\centering

\includegraphics[width=8.5cm,height=8.0cm,clip=true]
{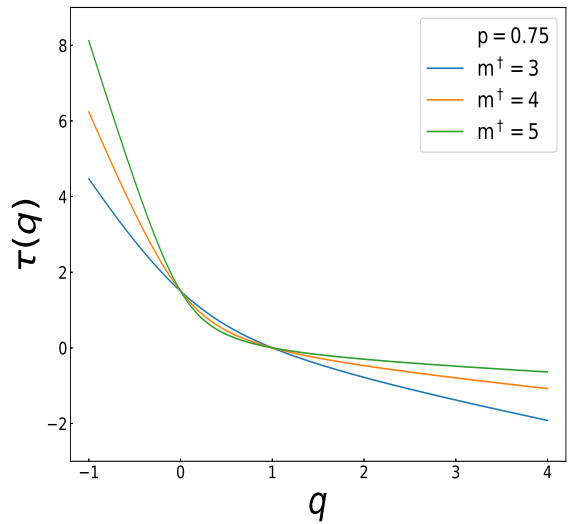}
\label{fig:tau}

\caption{ The plot of $q$ vs $\tau(q)$ for $p=0.75$ and $m^\dagger = 3, 4$ and $5$. 
All the plots intersects in two common points one of which represents $\tau(0)=0.5$ and 
another one represents $\tau(1)=0$ for any $m^\dagger$.
}

\label{fig:tau}
\end{figure}

Often we find that Legendre transform of a function can provide a better physical interpretation than 
the function itself. The Legendre transformation of $\tau(q)$ means we use its slope as an independent
parameter instead of $q$ itself. The slope of $\tau(q)$ is known as
Lipschitz-H\"{o}lder exponent $\alpha(q)$ and it is defined as
\begin{equation}
\label{eq:alpha_q}
    \alpha(q)=-{{d\tau(q)}\over{dq}},
\end{equation}
and hence we find
\begin{eqnarray}
\label{alpha_value}
\alpha(q) & = & \Big( m^\dagger+ {{m^\dagger b}\over{2m^\dagger-b}}-2\Big) -  \nonumber \\ & & \dfrac{q(m^\dagger-{{m^\dagger b}\over{2m^\dagger-b}})^2}{\sqrt{q^{2}(m^\dagger-{{m^\dagger b}\over{2m^\dagger-b}})^2+b^2}} .
\end{eqnarray} 
Using the definition of $\alpha(q)$ we find the Legendre transform of $\tau(q)$ which is 
\begin{equation}
\label{Legendre}
f(\alpha)=\tau(q) + \alpha q,
\end{equation} 
and hence the $f(\alpha)$ spectrum of the porous lattice is 
\begin{equation}
f(\alpha(q))=(p-1)+\dfrac{b^2}{\sqrt{q^{2}(m^\dagger-{{m^\dagger b}\over{2m^\dagger-b}})^2+b^2}}.
\end{equation}
It implies that each and every conservation laws obtained by tuning the $m^\dagger$ value results
in a spectrum of spatially intertwined fractal dimensions revealing the fact that the present 
system has infinitely many multifractal $f(\alpha)$ spectrum which share the same support. Note that $f(\alpha(q))$ is always concave in character (see figure 6) with a single maximum $2p$ at $q=0$ which corresponds to the dimension of the support. Since the dimension of the support $2p$ is always less than the dimension of the support $d = 2$ for $p<1$, so the support is always a fractal.

\begin{figure}
\centering

\includegraphics[width=8.5cm,height=8.0cm,clip=true]
{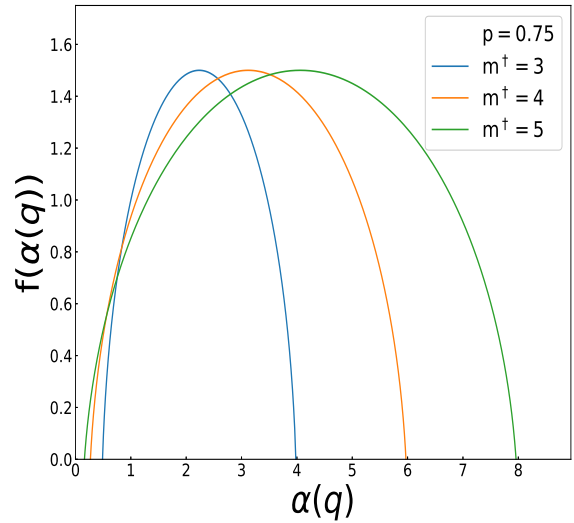}
\label{fig:7}

\caption{ The $f(\alpha)$ spectrum for $p=0.75$ and $m^\dagger = 3, 4$ and $5$. 
The maximum of each plot occurs at $q=0$ which is in fact the dimension $d_f$ of the support
$2p=1.5$. It confirms the theoretical prediction.}

\label{fig:7}
\end{figure}

\section{Area size distribution function and dynamic scaling in the stochastic porous lattice}

The snapshot of the lattice shown in Fig. (\ref{fig:2}) reveals the complex structure of the lattice in long time limit with a pattern of removed and surviving cells. Although the lattice is highly disordered, we have seen some order and non-trivial properties of the lattice in previous sections. It is then reasonable to ask whether the lattice grows with time in a self-similar fashion, that is, whether two or more snapshots of the lattice taken at different times are similar to each other or not. In Physics, 
there exist different kind of self-similarity. An object is called spatial self-similar 
if a suitably chosen part of the object is similar to the whole. The litmus test of this property is the
power-law behaviour of an observable. On the other hand, when a system evolves probabilistically
with time and the snapshots taken at different times are look-alike at least in the statistical
sense, then the system is said to exhibit temporal self-similarity. It means that the
 numerical values of the dimensional quantities are different in different snapshots but the corresponding dimensionless quantities remain the same. The litmus test of this property is the dynamic scaling \cite{ref.pavel, ref.mz, ref.khassan}

We can characterize each block by their area of the blocks in the lattice and investigate
the nature of their distribution. To this end, we define an observable quantity $C(a,t)da$ as 
the number of blocks whose area lies in the range $a$ and $a+da$ at time $t$. To calculate the distribution function we collect data for different times by finding the frequency of number of blocks 
using $\delta a$ as an interval size. The distribution function is then normalized by dividing it with the interval size $\delta a$. The total area of the remaining blocks is the given by $\int_0^\infty a C(a,t) da $ which according to Eq.(\ref{momenteq_ac}) gives,
\begin{equation}
\label{dc_1}
\int_0^\infty a C(a,t) da = \dfrac{1}{\Gamma(1+p)} t^{-(1-p)},
\end{equation}
and the number of remaining blocks is
\begin{equation}
  \int_0^\infty C(a,t) da =t^p, 
\end{equation}
which are essentially $M(2,2;4)$ and $M(1,1;t)$ respectively.

We shall now apply the Buckingham Pi theorem and invoke the idea of dimensional analysis to
check whether $C(a,t)$ exhibits dynamic scaling or not. The mean area of the remaining blocks $<a(t)>$ varies with time $t$ as $<a(t)> \sim t^{-1}$, as shown in Eq.(\ref{delta}). Now, the governed 
parameter $C$ depends on two governing parameters $a$ and $t$. But since $<a(t)> \sim t^{-1}$, it immediately implies that using $t^{-1}$ as a the yard-stick the areas of all the remaining blocks
can be measured. In other words, time $t$ be taken as an independent parameter as area $a$ can
be expressed in terms of $t$ and hence we can define a dimensionless governing parameter,
\begin{equation}
\label{xi}
\xi = \dfrac{a} {t^{-1}} = at. 
\end{equation}
It also means $C$ too can be expressed in terms of $t$ alone so that $C\sim t^\theta$
and therefore we can define a dimensionless governed parameter as
\begin{equation}
\label{Pi}
\Pi = \dfrac{C(a,t)}{t^{\theta}}.
\end{equation}

Now, according to the definition of dimensionless quantity, the numerical value of $\Pi$ will remain same even if the time $t$ is scaled by some factor. However $\Pi$ may still depend on the dimensionless governing parameter $\xi$ and hence we can write,
\begin{equation}
\label{Pi_sim}
\Pi \sim \phi(\xi).
\end{equation}
Using Eq. (\ref{Pi}) in the above equation
we find that the solution of $C(a,t)$ should have the following dynamic scaling form 
\begin{equation}
\label{disfunc}
C(a,t) \sim t^{\theta} \phi (at),
\end{equation}
where $\phi(\xi)$ is called the scaling function. The exponent $\theta$ can be calculated if we substitute Eq. (\ref{disfunc}) into Eq.(\ref{dc_1}), which gives,
\begin{equation}
\label{area_eq1}
\int_0^\infty a t^{\theta} \phi(at) da = \dfrac{1}{\Gamma(1+p)} t^{-(1-p)} ,
\end{equation}
which after change of variable becomes,
\begin{equation}
\label{area_eq2}
t^{\theta-2} \int_0^\infty  \xi \phi(\xi) d\xi = \dfrac{1}{\Gamma(1+p)} t^{-(1-p)}.
\end{equation}
The integration over the dimensionless quantity will result into another dimensionless quantity which immediately gives $\theta = 1+p$. So,
\begin{equation}
\label{Ct}
C(a,t) \sim t^{1+p} \phi(at).
\end{equation}
It implies that the plots of $C(a,t)$ versus $a$ for different time $t$ will be distinct. However, if the
same data is plotted for $C(a,t)t^{-(1+p)}$ as a function of $at$ they would collapse into a universal
scaling function since both $C(a,t)t^{-(1+p)}$ and $at$ are dimensionless quantities.

\begin{figure}
\centering
\label{fig:area_dc}
\includegraphics [width=8.5 cm,height=8 cm,clip=true]
{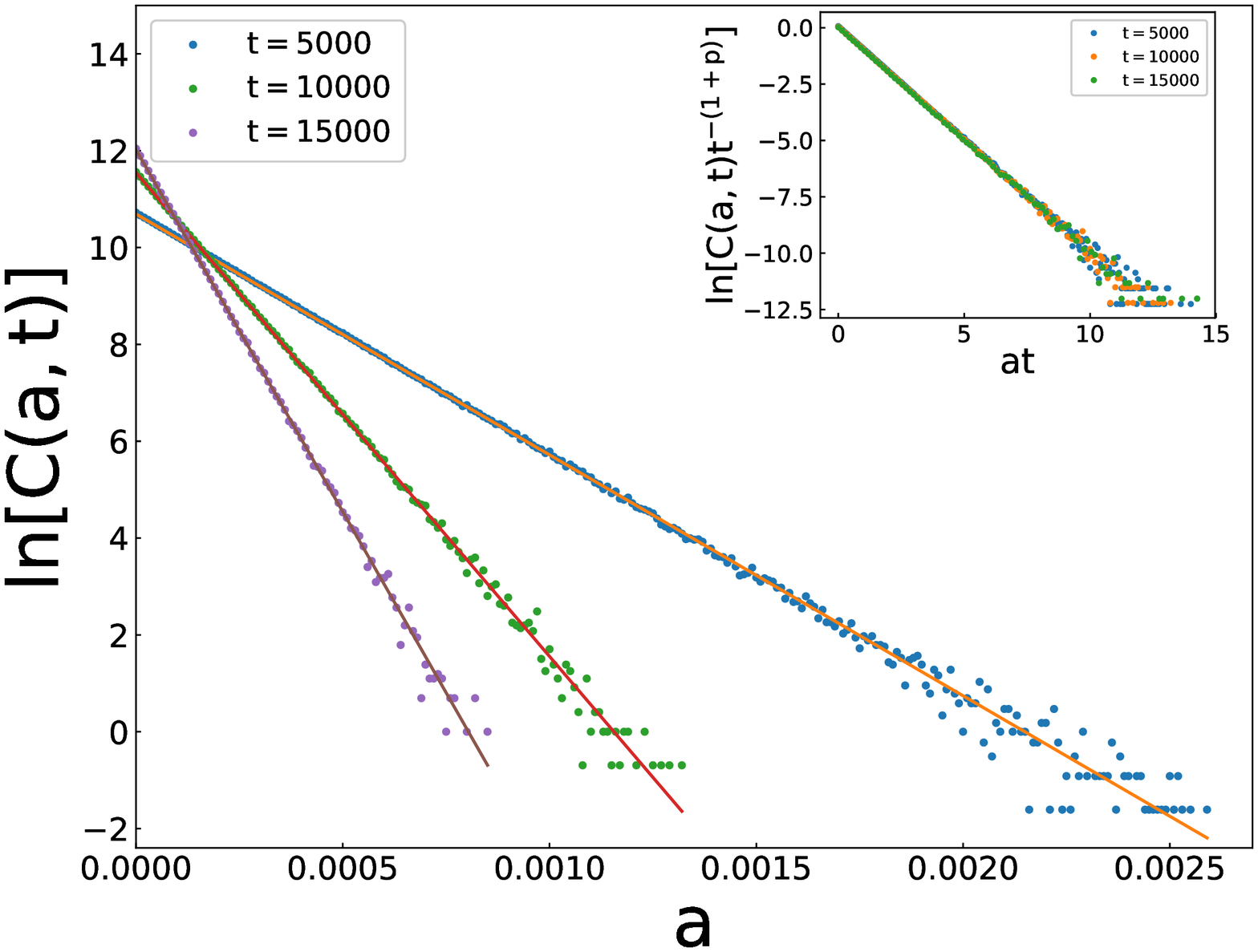}

\label{fig:area_dc}

\caption{The natural logarithm of the area distribution function $C(a,t)$ as a function of $a$ for $t=5000, 10000$ and $15000$ and $p=0.25$. The resulting plots are straight lines with slopes equal to $t$ and intercepts equal to $t^{1+p}$. The same graph is plotted in the inset of the figure where we have plotted $ln(C(a,t)t^{1+p})$ vs $at$ and it can be seen that all the distinct plots of the main plot collapses into a single universal curves.}

\label{fig:area_dc}
\end{figure}

We now examine the behavior of the area distribution function $C(a,t)$ to see how it varies with $a$. For this we have taken data for $p=0.25$ for $t=5000, 10000$ and $15000$ which are averaged over a large number of independent realizations. We have plotted $\ln[C(a,t)]$ as a function of $a$ for fixed times $t = 5000, 10000$ and $15000$ as shown in Fig. (\ref{fig:area_dc}). It is clear from the figure that the corresponding plots are distinct straight lines. The slopes of these straight lines were found to be equal to the time
$t=5000, 10000$ and $15000$ when the snapshots are taken and it confirms that,
\begin{equation}
\label{scaling}
C(a,t) \sim e^{-at}.
\end{equation} 
The intercepts were found to be equal $10.682,11.545$ and $12.047$ for $t=5000,10000$ and $15000$ respectively. These intercepts are equal to $ln(t^{1+p})$ for all the three values of $t$ which is in accordance with Eq. (\ref{Ct}). So, we can write that,
\begin{equation}
\label{area_final}
C(a,t) \sim t^{1+p} e^{-at},
\end{equation} 
and hence the solution for the scaling function is $\phi(\xi) \sim e^{-\xi}$.

Now, we examine whether the area distribution function possesses self-similarity using the idea of data collapse. To do this $C(a,t)$ is measured using $t^{(1+p)}$ as an yardstick and a is measured using $t^{-1}$ as an yardstick which makes them dimensionless governed quantity and dimensionless governing quantity respectively. As we have discussed earlier, the dimensionless quantities should remain the same for all $t$. We have plotted $\ln(C(a,t)t^{-(1+p)})$ vs at in the inset of Fig. (\ref{fig:area_dc}) and it can be seen that all the distinct plots of the main plot of $ln(C(a,t)t^{1+p})$ vs $at$ collapses into a single universal plot ignoring the fat tails. This universal plot now contains all the information about $C(a,t)$ in any time including the infinite time limit. This proves that although the numerical values of the dimensional quantities are different for each different time $t$, the corresponding dimensionless quantities remain invariant. So C(a,t) obeys dynamic scaling and the snapshots at different time $t$
are self-similar from the perspective of area size distribution.

\section{Summary and Conclusions}

In this article, we have formulated a bridge between stochastic dyadic Cantor set and weighted 
planar stochastic lattice with one crack (WPSL1). We have shown that if probabilistic mass loss 
is incorporated in WPSL1 such that a cell is kept with probability $p$ and removed with 
probability $1-p$, it can mimic a lattice with porous character. To the best of our
knowledge there do not exist a 
random lattice with coordination number disorder and have porosity of different sizes 
too. We have given a generalized master equation and solved it to find 
the various moments of the block size distribution functions. The trivial conserved 
quantity of WPSL1, that is the total mass is no longer conserved rather it decreases 
as $t^{-(1-p)}$. On the other hand, the total number of 
blocks increases with time as $t^{p}$ revealing that the mean area of the block size decreases
as $t^{-1}$. The numerical results matched perfectly and are consistent with the analytical 
results in all cases. 

The stochastic porous lattice also possesses infinitely many hidden 
conservation laws for each value of $p$ which are highly non-trivial 
since we can only know about them thanks to analytical solution. Moreover the ensemble 
average value of each of 
the conserved quantities was found to be equal to $1$ and it was 
also proven analytically albeit they are different from $1$ in every independent relization.
However, each of the infinitely many conservation laws is a 
multifractal measure, and so WPSL1 with mass loss possesses multiple multifractality. 
We have obtained an exact analytical generalized for the $f(\alpha)$ spectrum. It shows that
the peaks of $f(\alpha)$ spectrum is equal to $2p$ which is the support of the 
multifractal measure and it is fractal provided $0<p<1$. Finding an exact analytical expression
for $f(\alpha)$ is often formidable task.

We have checked whether the snapshot of the complex structure of the stochastic porous lattice 
taken at different times are similar or not. To this end, we have used the area distribution function and incorporated 
dynamic scaling and Buckingham Pi theorem. The scaling solution of the area distribution 
function $C(a,t)$ is found to exhibit dynamic scaling $C(a,t) \sim t^{1+p} e^{-at}$. To show
this we have used the idea of  
data-collapse. We have shown that the distinct plots of $C(a,t)$ as a function 
of $a$ for the snapshots at different time collapses superbly into on universal
curve if we plot $t^{-(1+p)}C(a,t)$ versus $at$ instead. 
It clearly suggests that the porous structure of the stochastic lattice 
possesses temporal self-similarity.

A two-dimensional counterpart of stochastic dyadic Cantor set leading to a porous structure 
has the potential of great use studying spreading of virus (computer or biological), fluids, opinion,
rumors etc. Besides, it can also be used as a skeleton for percolation process to see the impact of
$p$ that determines the degree of porosity. We can also see if percolation on it belongs to a 
new universality class or to the same class where WPSL belong to regardless of the $p$ value.
We intend to address some of these issues in our future endeavors.

\end{document}